\documentclass[times,twocolumn,final,authoryear]{elsarticle}
\usepackage{jasr}
\usepackage{framed,multirow}

\usepackage{amssymb}
\usepackage{latexsym}

\usepackage{url}
\usepackage{xcolor}
\definecolor{newcolor}{rgb}{.8,.349,.1}

\usepackage[citebordercolor=white]{hyperref}

\journal{Advances in Space Research}

\begin{document}

\verso{S. Komossa \textit{et al.}}

\begin{frontmatter}

\title{The extremes of AGN variability: outbursts, deep fades, changing looks, exceptional spectral states, and semi-periodicities} 

\author[1]{S. \snm{Komossa}\corref{cor1}}
\cortext[cor1]{Corresponding author: 
  skomossa@mpifr.de}
\author[2]{D. \snm{Grupe}}
\author[3]{P. \snm{Marziani}}
\author[4,5]{L. \v C. \snm{Popovi\'c}}
\author[4,5]{S. \snm{Mar\v ceta-Mandi\'c}}
\author[5]{E. \snm{Bon}}
\author[4,6]{D. \snm{Ili\'c}}
\author[4]{A. B. \snm{Kova\v cevi\'c}}
\author[1]{A. \snm{Kraus}}
\author[7,8]{Z. \snm{Haiman}}
\author[9,10]{V. \snm{Petrecca}}
\author[9,10,11]{D. \snm{De Cicco}}
\author[5]{M. S. \snm{Dimitrijevi\'c}}
\author[12]{V. A. \snm{Sre\'ckovi\'c}}
\author[5]{J. \snm{Kova\v cevi\'c Doj\v cinovi\'c}}
\author[13]{M. \snm{Pannikkote}}
\author[5]{N. \snm{Bon}}
\author[14,15]{K. K. \snm{Gupta}}
\author[16]{F. \snm{Iacob}}

\address[1]{Max-Planck-Institut f\"ur Radioastronomie, Auf dem H\"ugel 69, 53121 Bonn, Germany}

\address[2]{Department of Physics, Geology, and Engineering Technology, Northern Kentucky University, 1 Nunn Drive, Highland Heights, KY 41099, USA}

\address[3]{National Institute for Astrophysics (INAF), Astronomical Observatory of Padua, Vicolo dell’Osservatorio 5, Padua IT, 35122, Italy}

\address[4]{University of Belgrade - Faculty of Mathematics, Department of Astronomy, Studentski trg 16, 11000 Belgrade, Serbia}



\address[5]{Astronomical Observatory, Volgina 7, 11000 Belgrade, Serbia}



\address[6]{Hamburger Sternwarte, Universit\"at Hamburg, Gojenbergsweg 112, 21029 Hamburg, Germany}



\address[7]{Department of Astronomy, Columbia University, New York, NY 10027, USA}

\address[8]{Department of Physics, Columbia University, New York, NY 10027, USA}

\address[9]{Dipartimento di Fisica "Ettore Pancini", Università di Napoli Federico II, via Cinthia 9, 80126 Napoli, Italy}

\address[10]{INAF - Osservatorio Astronomico di Capodimonte, Salita Moiariello 16, 80131 Napoli, Italy}



\address[11]{Millennium Institute of Astrophysics, Nuncio Monseñor Sótero Sanz 100, Of 104, Providencia, Santiago, Chile}


\address[12]{Institute of Physics, University of Belgrade, Pregrevica 118, 11080 Belgrade, Serbia} 


\address[13]{Physics Department, Tor Vergata University of Rome, Via della Ricerca Scientifica 1, 00133 Rome, Italy}


\address[14]{STAR Institute, University of Liège, Quartier Agora, Allée du Six Auôt 19c, 4000, Liège, Belgium}

\address[15]{Sterrenkundig Observatorium, Universiteit Gent, Krijgslaan 281 S9, 9000, Gent, Belgium}

\address[16]{West University of Timisoara, 300232 Parvan Blvd. no. 4, Timisoara, Timis, Romania}

\received{11 July 2024}
\finalform{X X 2024}
\accepted{X X 2024}
\availableonline{2024}
\communicated{S. Sarkar}

\begin{abstract}
The extremes of Active Galactic Nuclei (AGN) variability offer valuable new insights into the drivers and physics of AGN.  We discuss some of the most extreme cases of AGN variability; the highest amplitudes, deep minima states, extreme spectral states, Seyfert-type changes, and semi-periodic signals, including new X-ray observations. 
The properties of changing-look (CL) AGN are briefly reviewed and a classification scheme is proposed which encompasses the variety of CL phenomena; distinguishing slow and fast events, repeat events, and frozen-look AGN which do not show any emission-line response. 
Long-term light curves that are densely covered over multiple years, along with follow-up spectroscopy, 
are utilized to gain insight into the
underlying variability mechanisms including accretion disk and broad-line region physics.
Remarkable differences are seen, for instance, in the optical spectral response to
extreme outbursts, implying distinct intrinsic variability mechanisms.
Furthermore, we discuss methods for distinguishing between CL AGN and CL look-alike events (tidal disruption events or supernovae in dense media). Finally, semi-periodic light curve  variability is addressed and the latest multiwavelength (MWL) light curve of
the binary supermassive black hole (SMBH) candidate OJ 287 from the MOMO project is presented. 
Recent results from that project have clearly established the need for new binary SMBH modelling
matching the tight new constraints from observations, including the measurement of a low (primary) SMBH mass of $\sim 10^8$ M$_\odot$ which also implies that OJ 287 is no longer in the regime of near-future pulsar timing arrays. 
\end{abstract}


\end{frontmatter}


\section{Introduction}
\label{sec1}


Active Galactic Nuclei (AGN) are thought to be powered by accretion of matter onto the supermassive black holes (SMBHs) at their centers. The central region is surrounded by two line-emitting systems, the broad-line region (BLR) closer to the nucleus and of higher density, and the narrow-line region (NLR) at larger distances up to kiloparsecs and of lower density \citep{Osterbrock1989}. 
The unified model of AGN has been successful in explaining the major differences between type 1 and type 2 AGN by viewing angle differences into the central engine \citep{Antonucci1993}. In addition, other parameters are known to play a role in explaining variations across the Seyfert and quasar populations and within each population, including the influence of the host galaxy, evolutionary effects, accretion rate, and/or metal abundances, for instance. BLR and NLR emission-line properties as well as continuum (spectral energy distribution; SED) shape have been used to evaluate differences across the AGN population using Principal Component Analyses \citep[e.g.,][]{Boroson1992, Sulentic2000a, Grupe2004, Xu2007, Marziani2018}, establishing the accretion rate as a main driver. Another approach involves the  investigation of continuum and emission-line variability. This is the topic of this contribution.  

The extremes of AGN variability provide valuable insights into the underlying drivers and physics of the central engine. 
In X-rays, extreme flux and spectral states can reveal the nature of the inner accretion disk, probe the physics of matter under strong gravity, may offer a way of measuring SMBH spin, and can uncover the mechanisms responsible for the expulsion of matter from the SMBH in form of strong outflows. 
Amplitudes of X-ray variability exceeding a factor 30-50 are still rare, and only a few systems are known to exceed a factor of 100 \citep[e.g., IC 3599, WPVS007, and 1ES\,1927+654;][]{Grupe2015, Grupe1995-wpvs, Ricci2020}.  
The highest amplitudes of X-ray variability to date from the cores of galaxies, exceeding factors of 1000, have nearly exclusively been observed in {\em non-active} galaxies in the form of temporary flares from stars tidally disrupted by dormant SMBHs (stellar tidal disruption events; TDEs). These latter will not be discussed much further in this contribution \citep[see][for a review]{Komossa2015}. 

In the optical regime, changing-look AGN (CL AGN hereafter) with highly variable emission lines which change the spectroscopic Seyfert type, provide us with tight new constraints on the physics of the outer accretion disk, the BLR and the coronal-line region (CLR). 
The term CL AGN was initially coined to describe systems which changed their X-ray spectrum from mildly or unabsorbed to heavily absorbed or Compton-thick  \citep{matt2003}. However, in recent years the term has been widely used to refer to optical spectroscopic Seyfert-type changes, from type 1 to type 2 (or 1.8, 1.9) and vice versa \citep[e.g.,][]{LaMassa2015}. We use this latter definition throughout this publication. 

\begin{figure}[h]
\includegraphics[clip, width=4.7cm]{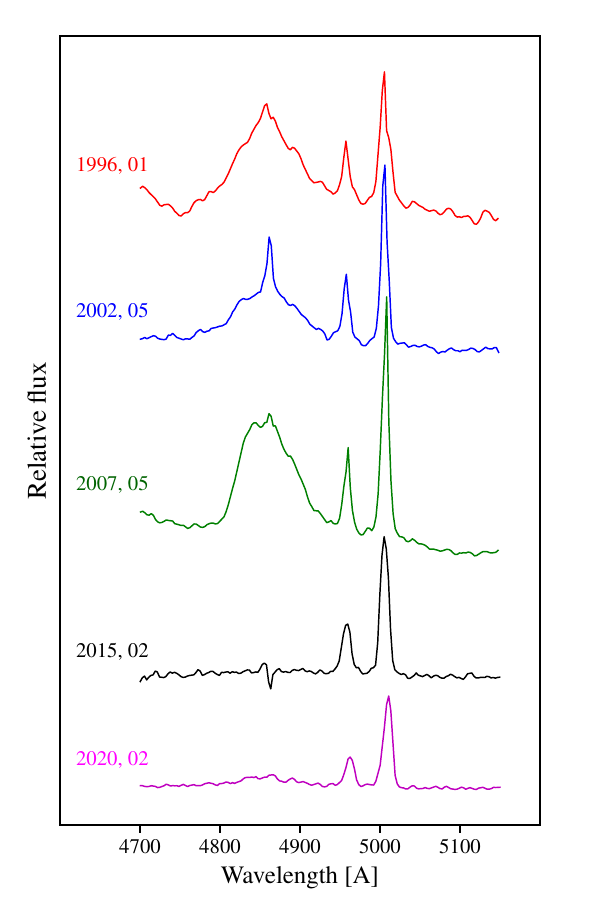}
\includegraphics[clip, width=8.7cm]{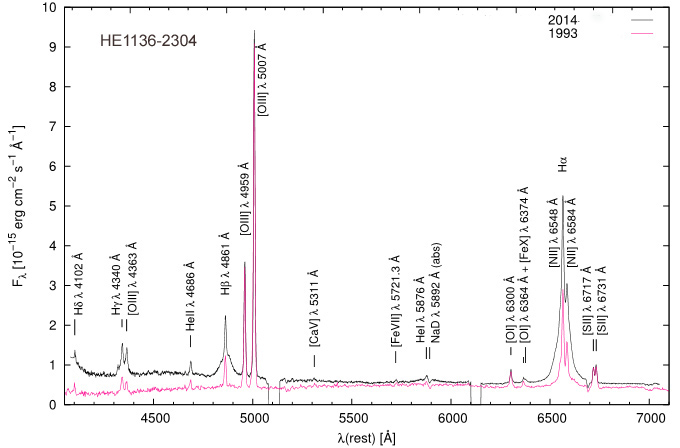}
   \caption{Selected optical spectra of CL AGN. Left: Selected H$\beta$--[OIII] spectra of NGC 3516 demonstrating its strong spectral changes from Seyfert type 1 to type 1.9, adapted from \citet{Popovic2023}. Right: Broad-band spectrum of
   HE 1136--2304 \citep[adapted from][]{Parker2016} first detected by its high-amplitude X-ray outburst. Follow-up optical spectroscopy then revealed its CL. 
}
\label{fig:CL-example-spectra}

    \end{figure}

Single cases of CL AGN were noticed already decades ago. One prominent example is NGC 4151.  
Traditionally known as intermediate-type Seyfert galaxy \citep{Seyfert1943}, its broad Balmer lines became very faint in 1984 \citep{Penston1984, Lyuty1984}, barely detectable but still
present \citep{Penston1984, Kielkopf1985}. They were already back to their bright state in January 1985 \citep{Peterson1985}. 
Two recent examples are shown in Fig. \ref{fig:CL-example-spectra}, initially selected either through their high-amplitude continuum flaring and the CL event was then found in triggered follow-up optical spectroscopy (HE 1136--2304;  results in Fig. \ref{fig:CL-example-spectra} taken from \citet{Parker2016}), 
or targeted in the course of reverberation mapping and long-term spectroscopy over decades 
(NGC 3516; results in Fig. \ref{fig:CL-example-spectra} taken from \citet{Popovic2023}).  While HE 1136--2304  only changed its Seyfert type from 1.5 to 1.8, NGC 3516 changed from type 1 to type 1.9. 

The topic of CL AGN  received particular attention in recent years, because larger samples were selected from optical surveys, more extreme cases were identified, events were detected not only in Seyfert galaxies but also in quasars where the emitting regions are much more extended so that longer variability timescales are expected, and higher amplitudes of variability were detected not only in X-rays but also in the optical band \citep[see][for recent reviews]{Ricci2022, KomossaGrupe2023}. 
The rapid and high-amplitude changes pose challenges to accretion-disk models \citep{Lawrence2018}, and many new scenarios for rapid changes in the disk emission have been developed in recent years. 

The first part of this review will focus on extreme cases of variability of radio-quiet AGN and the second part will briefly address semi-periodic variability of candidate binary SMBHs.  In Sect. 2 we review the frequency of CL AGN. In Sect. 3 we present a new scheme which sorts the CL AGN into different phenomenological categories based on their continuum and line variability. 
Sect. 4 briefly summarizes the different theoretical approaches suggested to understand CL AGN, and in Sect. 5 representative  cases of extreme flux and spectral variability are discussed, including new X-ray observations. Sect. 6 provides a critical summary of observational effects which can lead to false CL detection and elaborates on how to distinguish true CL events from CL look-alikes. 
Sect. 7 addresses (semi-)periodic variability of radio-loud systems, which has been suggested to be associated with the presence of binary SMBHs. A focus is on the blazar OJ 287. The article concludes with an outlook into future time domain surveys (Sect. 8).
Throughout this publication, we use a cosmology with 
$H_{\rm 0}$=70 km\,s$^{-1}$\,Mpc$^{-1}$, $\Omega_{\rm M}$=0.3 and 
$\Omega_{\Lambda}$=0.7
and the cosmology calculator of \citet{Wright2006}. 

\section{The frequency of CL AGN}

Large-area spectroscopic surveys offer excellent data bases for selecting samples of CL AGN. Ultimately, these will provide us with the frequency, timescales, and disk and host properties of CL events in Seyfert galaxies and quasars.
A number of selection effects have to be kept in mind: For instance, if the starting sample consists of type 1 AGN, then turn-offs (decreases in line and continuum luminosity) will be preferentially found in the follow-up spectroscopy. If the starting sample consists only of type 2 AGN, then only turn-ons, if any, will be found at the second epoch of spectroscopy. 
Further, the amplitude of the type change (into type 1.8, 1.9, or 2) will depend on telescope sensitivity, with faint Balmer lines more difficult to detect in spectra of lower signal-to-noise ratio (see also Sect. 6). 
Further, AGN with already faint Balmer lines (lower Eddington accretion rates at fixed SMBH mass) will be preferentially detected as CL AGN (as turn-offs), because only a small change in the luminosity is needed to make the lines undetectable, and therefore make the AGN appear as CL. If, instead the broad lines are luminous, a much larger luminosity change is required to identify a CL event. An alternative could be therefore to require a minimum factor (10) of variability, independent of exact AGN type, to search for CL AGN. 
Finally, the selection criteria used to identify CL AGN (e.g., the required amplitude of the type change) so far have varied between different studies which needs to be taken into account as well when directly comparing results. 

Larger-sample searches for CL AGN have been carried out recently with, for instance, LAMOST \citep{Yang2018}, 6dFGS/ Skymapper \citep{Hon2022}, and DESI \citep{Guo2023}. Otherwise, mostly 
the exceptional data base of the Sloan Digital Sky Survey \citep[SDSS;][]{York2000} has been employed
for subselecting larger samples of AGN which have been observed at least twice with SDSS or once with SDSS and followed-up at other telescopes \citep[e.g.,][]{Runco2016, Ruan2016, MacLeod2016, Wang2018, MacLeod2019, Graham2020, PottsVillforth2021, Green2022, Zeltyn2024}. 
These have started to quantify the frequency and properties of CL events and types in Seyfert galaxies and quasars. 

\citet{Runco2016} found that the frequency of Seyfert galaxies significantly changing their spectroscopic type was 10\% in a sample of nearby type 1 Seyferts initially selected from SDSS and then followed up with Keck. The same frequency in an SDSS selected sample of quasars was lower at $\sim$1\% \citep{MacLeod2016}. Approximately equal numbers of events got brighter and changed towards type 1, or got fainter and changed towards type 2. 
The largest sample of CL AGN selected recently from SDSS V \citep{Zeltyn2024} contained 67\% CL AGN which showed  dimming and 33\% which showed brightening. The CL AGN have a median accretion rate of $L/L_{\rm{Edd}}$ = 0.025, slightly lower than in a control sample with $L/L_{\rm{Edd}}$ = 0.043.    

Archival searches on large spectroscopic samples are excellent for statistical inferences on the CL AGN population. An alternative approach has been to carry out rapid MWL follow-up observations within days to weeks, once new CL AGN candidates were identified by the onset of flaring or dipping in the optical or X-ray band \citep[e.g.,][]{Parker2016}. That way, dense monitoring of the ongoing, evolving event is possible. Representatives of such AGN are discussed in Sect. 5.

\section{Definition of different categories of CL AGN}

In recent years, many different cases of high-amplitude BLR and continuum variability have been reported, making CL AGN quite a mixed bag, with possibly a variety of different underlying physical processes operating in different systems and at different timescales. 
Therefore, here, we first introduce several different phenomenological CL AGN categories, based on the type of continuum variability and/or line response: 

$\bullet$ Category I: Slow switch-on AGN. These are CL AGN which slowly turn on, on a timescale of years to decades.  An example for this pattern may be AT2017bcc \citep{Ridley2024}, even though the event is unusual and was initially interpreted as superluminous supernova or a candidate tidal disruption event (see also our Sect. 6). 

$\bullet$ Category II: Slow switch-off AGN. 
In this category are CL AGN which slowly turn off, on a timescale of years to decades.
A prime example for this category is Mrk 590. Once known as a classic Seyfert 1 galaxy, its optical continuum emission diminished by a factor of 100 over two decades \citep{Denney2014}. In low-state, the only broad emission line still visible in the optical spectrum is a weak component of H$\alpha$, implying a change from type 1 to type 1.9. 

$\bullet$ Category III: Rapid outbursts. In this category are CL AGN which spend most of their time in fainter states but then show a rapid (timescales of weeks to months) strong increase in their optical--X-ray continuum luminosity, followed by a response in the broad emission lines.  Prime examples are IC 3599 \citep{Grupe2015} and NGC 1566 \citep{Parker2019} which showed dramatic increases of their continuum flux on a timescale of weeks to months.  

$\bullet$ Category IV: Rapid drops. In this category are CL AGN which spend most of their time in brighter states but then show a  rapid (timescales of weeks to months) decrease in their optical--X-ray continuum  and broad-line flux. 

$\bullet$ Category V: Regular, repeating brightening and dimming events. In this category are CL AGN which show recurrent medium-amplitude increases and decreases in their line and continuum brightness. Many of the reverberation-mapped AGN are in this category, for instance NGC 5548 \citep{Bon2016}. 

$\bullet$ Category VI: Frozen-look AGN.  
In this category are AGN which do not show any emission-line changes in response to strong continuum variability. 
This category is rare. Examples include 
saturation effects in the BLR ionization in case of particularly high continuum luminosity \citep{Pronik1972, Pei2017, Panda2023}.
A much stronger example is the lack of HeII response despite high-amplitude soft X-ray continuum variability of Mrk 335 \citep{Komossa2020-m335}, explained in this case by changes of (dust-free) absorption occurring only along our line-of-sight, not seen in other directions (i.e., not seen by the bulk of the emission-line gas).    

\begin{table*}[t]
\centering
\caption{Classification of CL-AGN according to the type of continuum variability and emission-line  response. All categories may repeat.} \label{tab:class}
\begin{tabular}{|l|l|l|}
\hline
category & CL classification & timescale \\
\hline
I & slow switch-on & years--decades  \\
II & slow switch-off & years--decades  \\
III & rapid transient event: outburst & rise time: weeks to months  \\
IV & rapid transient event: deep fade & fade time: weeks to months  \\
V & regular brightenings and dimmings & weeks to years  \\
VI & frozen-look events & any time \\
\hline
\end{tabular}
\end{table*}

Several CL events are known to repeat. Repetition may occur in all categories, but it depends on the cadence and duration of the observational coverage how easily recurrence can be detected. Single AGN may even cross categories in the long term.   
Some processes are more common than others. 
Tab. \ref{tab:class} provides a summary of the different categories of CL AGN. 
The strongest cases of CL AGN are in categories I--IV since they exhibit the largest and most unexpected changes and place most tight constraints on variability models. 

Finally, we note that CL AGN were first identified based on observations, and therefore the proposed categories closely match what has been observed so far. They do not yet include any underlying theoretical scenarios, because it is still unknown how to explain CL AGN, and many different scenarios have been proposed in the literature. Not all of them make predictions for the timing and spectral evolution, yet. 
In the next Section, we provide a short overview over the different models for CL AGN  which have been proposed in recent years. 

\section{(New) models for extreme AGN variability}

There are two fundamentally different ways to explain high-amplitude variability. One of them is intrinsic variability where the emission of the accretion disk and BLR itself changes.
The other is apparent variability where the intrinsic emission remains constant, but is then affected on its way to the observer.  Such extrinsic variability can be caused by the effects of gas absorption and dust extinction, or by the effect of gravitational lensing.
We discuss the different mechanisms in turn.

\subsection{Extrinsic variability: gas absorption and dust extinction}

Gas absorption affects the X-ray continuum emission through Hydrogen and metal K-shell absorption. For gas column densities exceeding  $\sim$10$^{23-24}$ cm$^{-2}$, the material becomes Compton-thick and no X-rays are detected at all below 10 keV where most X-ray satellites operate.
If an absorber contains dust, it also significantly affects the optical continuum through dust extinction, causing a characteristic reddening of the continuum shape and a deviation of the BLR Balmer decrement from its (case B) recombination value.

Dust extinction and/or gas absorption is therefore an efficient mechanism to change the observed luminosity of an AGN, and if at work the mechanism can be relatively easily identified  through its characteristic effects on the optical and X-ray spectra.
However, while partial obscuration from outflowing dusty clouds in the outer BLR can significantly affect line profile shapes \citep{GaskellHarrington2018}, and while single CL AGN do show effects of significant extinction \citep{PottsVillforth2021}, this mechanism has generally not been favored for the majority of CL systems, because the expected BLR reddening was not observed \citep{Alloin1985}, the optical-UV continuum shape was inconsistent with extinction \citep{MacLeod2016}, the associated X-ray absorption was not observed \citep{Parker2019}, the timescale of variability was too fast \citep{LaMassa2015}, and/or the level of polarization was very low \citep{Hutsemekers2019}. 

\subsection{Extrinsic variability: gravitational lensing}

Lensing or microlensing can magnify (or demagnify) the accretion disk and/or the BLR emission.  Depending on the geometry, it can give rise to achromatic light curve variability and/or it can affect different parts of the emitting region, causing complex line profile shapes and changes.
Microlensing modelling of BLR profile variations has not yet been much applied to changing-look AGN, but studied in a number of known lensed quasars \citep{Fian2021, Savic2024}. It is rare and is therefore not expected to explain the population of nearby CL AGN as a whole.

\subsection{Intrinsic variability: Modified lampposts, disk instabilities, disk precession, influence of magnetic fields, BLR stability, binaries}

CL events challenge our ideas of the properties of the accretion disk in AGN. 
Given the rapid and high-amplitude variability of some CL AGN, a large number of new theoretical studies in the last few years have explored new mechanisms for accretion disk structural and emission changes. Some of these addressed 
mechanisms of a more rapid transfer of changes across the disk (e.g., involving magnetic fields), but left open the origin of the high-amplitude outburst in the first place; others addressed the outburst mechanism directly; and yet others explored changes of the BLR structure itself.

First, a modified lamppost model has been suggested \citep{Lawrence2018}, assuming that the X-ray emission which arises from the lamppost is not directly reprocessed by the accretion disk, but rather by a system of dense gas clumps above the disk. This explains a rapid MWL response, but leaves open the mechanism(s) of the intrinsic high-amplitude of the X-ray variability itself. 

Second, different types of accretion disk instabilities have been considered to explain high-amplitude continuum changes, including the radiation-pressure instability of the inner disk \citep{Lightman1974, Grupe2015, Sniegowska2020}, a Hydrogen ionization instability \citep{NodaDone2018}, or instabilities in warped disks at large radii \citep{RajNixon2021}. An instability in the circumnuclear gas supply has been discussed as well 
\citep{Wang2024}. Tidal effects on the mini-disks surrounding both SMBHs of a binary SMBH system \citep{WangBon2020} have seen suggested as well.  

Third, the influence of magnetic fields, or a combination of several effects, has been considered in different ways \citep{Ross2018, Stern2018, DexterBegelman2019, Scepi2021, PanXin2021, Laha2022, Cao2023}. 

Fourth, effects of two types of shocks developing in highly tilted, highly precessing disk were discussed \citep{Kaaz2023}. 

Finally, some models were suggested where the BLR gas 
itself is directly affected, through an instability \citep{Nicastro2000}, initially proposed to explain the absence of a BLR in low-accretion-rate systems.  

\subsection{New approaches of understanding BLR and CLR physics}

Independent of the exact mechanism which causes the change in the illumination of the emission-line gas, the reprocessing of the radiation on BLR and CLR spatial scales provides us with a wealth of new applications regarding BLR and CLR physics, and the underlying atomic processes. 
We just give a few examples here. For instance,
some of the highest-amplitude outbursts are known to produce a strong response not only in the BLR, but also in the CLR. Strong high-ionization lines in the CLR temporary excited by an outburst provide tight constraints on the intrinsic, unobservable EUV shape of the SED because their ionization potentials are located in the EUV to soft X-ray regime. Further, atomic parameters of many coronal lines are still poorly known with uncertainties in atomic parameters and collision strengths as large as a factor of 10 \citep[e.g.,][]{Mohan1994, Pelan1995, Oliva1997, Berrington2001}
given the difficulties of producing them in laboratory plasmas. In addition, rare coronal line transitions can be excited that way, and their ratios provide us with valuable new diagnostics of the density and temperature \citep{NussbaumerStorey1982} 
of the CLR in AGN.  

A second example regards the BLR physics. The nature of the outer BLR is still not well understood. Once gas temperatures drop sufficiently below $\sim$2000 K  the gas conditions change significantly, Hydrogen molecules can be present \citep{Crosas1993}, and new ionization/recombination processes start to dominate (see \citet{Sreckovic2018, Sreckovic2020, Milan2021} for a detailed discussion of the relevant physical processes, including chemi-ionization/recombination processes, the importance of H$^*$(n) + H(1s) collisional ionization, the influence of inverse recombination processes, and 
in \citet{Milan2021}
the  importance of excitation/de-excitation processes with (n-n') mixing in such collisions).  Turn-off CL AGN may be ideally suited to probe the presence and physics of such an outer BLR component, as the inner parts of the BLR will fade first in response to a deep drop in the ionizing continuum luminosity, systematically leaving the outer parts of the BLR detectable in high-cadence, time-resolved spectroscopy of the BLR response.  

\section{Extreme cases of AGN variability and their interpretation}

Given the different types of extreme continuum and spectral variability of AGN and the relatively low number of sources, we discuss these on an individual basis, focussing on systems for which we have obtained long-term observations, and adding new data from the Swift archive as well. The object properties are listed in Tab. \ref{tab:basic-prop}.


\begin{table*}[h]
\centering
\caption{Properties of the highly variable AGN discussed in this contribution. Column entries are as follows: (1) galaxy name, (2) redshift, (3) spectroscopic type change, (4) total amplitude of variability in X-rays $A_{\rm X, var}$ detected with Swift,  and (5) comments. The X-ray amplitude for NGC 5548 is based on the second-lowest data point, since the lowest is an outlier. }
\label{tab:basic-prop}
\begin{tabular}{|l|c|l|c|l|}
\hline
name & $z$ &  Seyfert-type change & $A_{\rm X, var}$ & comments\\
(1) & (2) & (3) & (4) & (5)  \\
\hline
IC 3599 & 0.020758  & 1 --> 1.9  & 202 &   \\
NGC 1566 & 0.005017  & 1 <--> 2  & 76 &  recurrent (strong) CLs \\
NGC 5548 & 0.017175 & 1.5 <--> 1.8  & 10 & recurrent (mild) CLs  \\
NGC 3516 & 0.008836 &  1 <--> 1.9  & 76 &  recurrent (mild) CLs \\
Mrk 335 & 0.025785 & --  & 119 &  \\
\hline
\end{tabular}
\end{table*}

\subsection{High-amplitude outbursts and the cases of IC 3599 and NGC 1566}

\subsubsection{IC 3599}

The Seyfert galaxy IC 3599 underwent a giant outburst detected in the course of the ROSAT all-sky survey. Its X-ray flux increased by more than a factor 100. Optical spectra taken by chance during the ongoing outburst revealed bright broad Balmer lines and high-ionization iron coronal lines \citep{Brandt1995}. These emission lines then faded strongly in subsequent years \citep{Grupe1995, KomossaBade1999}. In low-state, among the Balmer lines, only broad H$\alpha$ is still significantly detected, leading to a Seyfert 1.9 classification \citep{KomossaBade1999, Grupe2015}. Faint coronal lines are still present in low-state. Photoionization modelling of the optical outburst spectrum revealed that the high-ionization lines are well explained with gas typical of a CLR, with a gas density of $\sim$10$^9$ cm$^{-3}$
\citep{KomossaBade1999}. 
The cause of the CL event of IC 3599 was not discussed in more detail at that time, but Narrow-line Seyfert 1 (NLS1) variability, a TDE, or an accretion-disk instability were briefly mentioned as possibilities; the latter favored by \citet{KomossaBade1999}. 

\begin{figure}[h]
 \includegraphics[clip, trim=0.3cm 5.3cm 1.7cm 11.0cm, width=8.7cm]{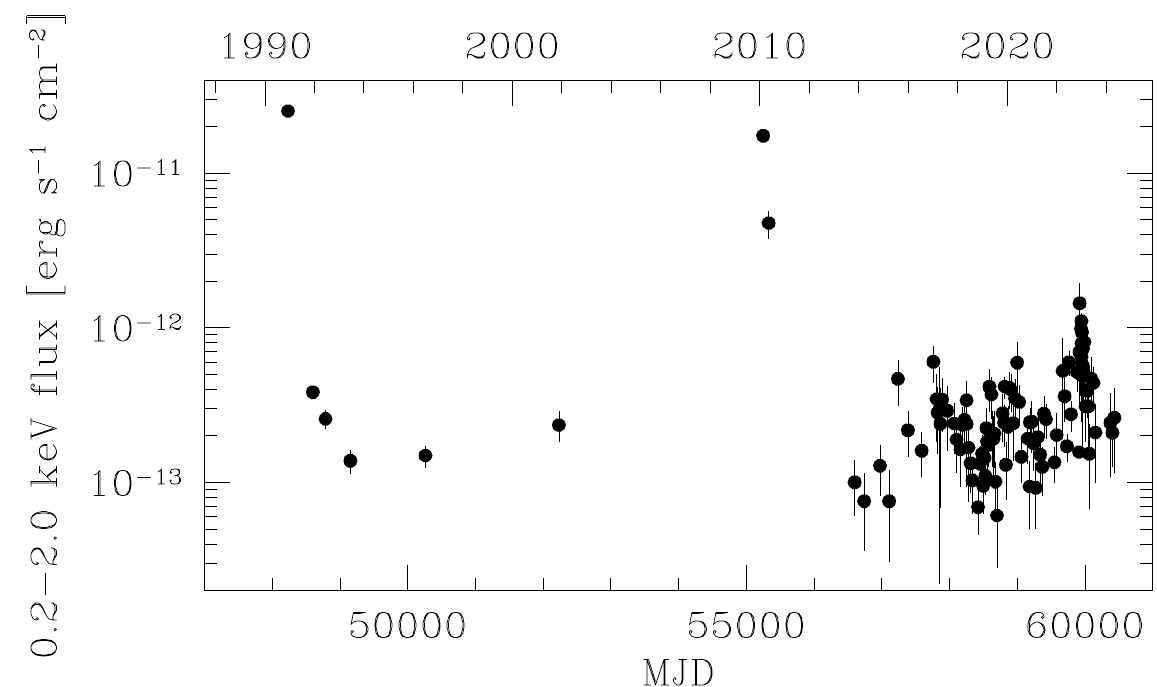}
   \caption{Long-term X-ray light curve of IC 3599 until 2024, with data from ROSAT, Chandra, and Swift. Two bright outbursts are visible in 1990 and 2010 with an amplitude of a factor $>$100.
}
\label{fig:lc_swift3599}%
    \end{figure}

\begin{figure}
 \includegraphics[clip, trim=0.9cm 1.1cm 1.9cm 0.5cm, angle=-90, width=8.7cm]{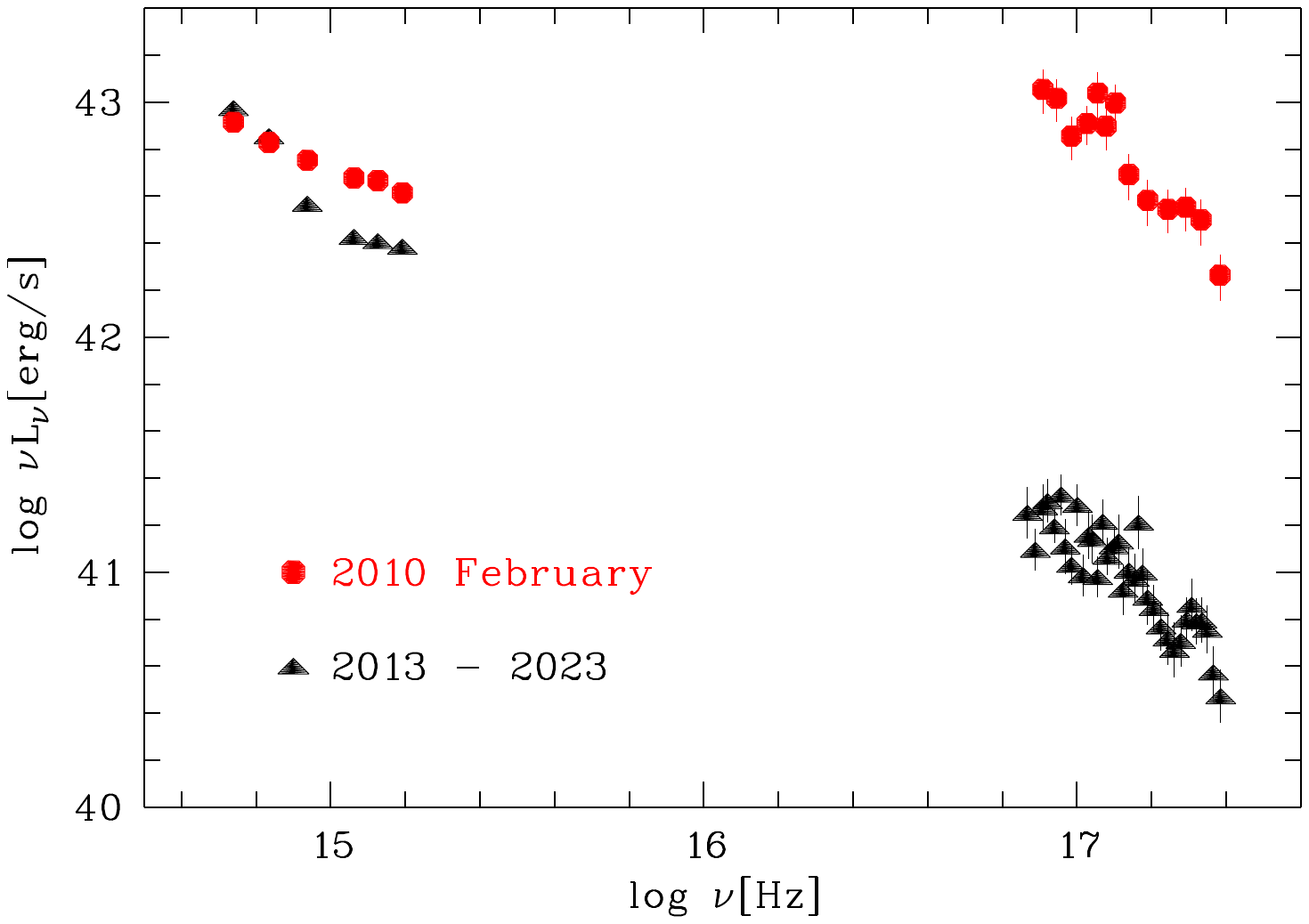}
   \caption{SED of IC 3599 \citep{Grupe2024} during outburst (red; in February 2010) and during quiescence (black; Swift low-state data points between 2013 and 2023 were co-added).
The X-ray spectrum is extremely soft, both in outburst and quiescence. A giant EUV--soft X-ray excess is present during outburst. }
\label{fig:SED_IC3599}%
    \end{figure}

In ongoing Swift monitoring, a second X-ray outburst of similar amplitude was detected in 2010, 19.5 yr after the first one \citep{Komossa2014, Campana2015, Grupe2015}. The outbursts are consistent with a Lightman-Eardley disk instability, using the observed flare repeat time of 19.5 yrs 
and a SMBH mass range of 10$^{6-7}$ M$_{\odot}$ inferred from observations, 
which then implies a truncation radius between inner and outer disk of $r_{\rm trunc}$ = 5--45 $r_{\rm g}$ \citep{Grupe2015}.  
MWL monitoring is ongoing in order to search for new high-amplitude outbursts. No third one has been detected so far \citep{Grupe2024}, but significant lower-amplitude X-ray variability is ongoing (Fig. \ref{fig:lc_swift3599}). During outbursts, the X-ray spectrum of IC 3599 is exceptionally steep ($\Gamma_{\rm x} \sim 4$). The cause for the steepness remains unknown, but it seems to point towards the absence of a disk corona. Even in low-states, the X-ray spectrum remains very soft (Fig. \ref{fig:SED_IC3599}). 
Our Swift monitoring of this extreme optical CL AGN is ongoing. Rapid MWL follow-ups including optical spectroscopy at dense cadence will provide us with a unique method of reverberation-mapping the BLR and CLR following the next giant outburst.

\subsubsection{Recurrent changing-look activity of NGC 1566} 

NGC 1566 is similar in several respects to IC 3599 including the detection of strong iron coronal lines during outburst, but with a lower amplitude in X-rays, particularly hard (not soft) X-ray spectra, and better optical spectroscopic coverage, so that one can tell that it went through repeated CL events. 

NGC 1566 is a very nearby NLS1 galaxy \citep{Xu2024} at redshift $z=0.005$. The widths of the broad Balmer lines of NGC 1566 are FWHM(H$\beta$) = 1950 km/s and FWHM(H$\alpha$) = 1970 km/s \citep{DaSilva2017, Alloin1985}. 
Fe\,II emission is detected in the type 1 and 1.5 spectral state \citep{Ochmann2024}.
NGC 1566 was known as a bright type 1 AGN in the 1960s \citep{Shobbrook1966}, but became much fainter in subsequent years.
Spectroscopic observations taken between 1980 and 1982 found it to be a type 2 in 1980, then gradually changing back to a type 1. Another CL event was detected in 1985 \citep{Alloin1985, Alloin1986}. Most of the changes occurred rapidly, within only 4 months. The BLR Balmer decrement did not change significantly during the CL event \citep{Alloin1985}.

The most recent outburst of NGC 1566 in 2018 was first detected with INTEGRAL in the hard X-ray band \citep{Ducci2018}. MWL follow-up observations were rapidly triggered \citep{Parker2019, Oknyansky2019, Ochmann2020} which resulted in the detection of a new CL event. 
During the outburst, the optical spectrum changed from type 1.8 to type 1, accompanied by strong variations in the helium lines and iron coronal lines \citep{Oknyansky2019, Ochmann2024}. 
The FWHM of the Balmer lines remains relatively constant during all observations. The emission lines OI$\lambda$8446 and CaII$\lambda$$\lambda$8498,8542,8662 are double-peaked \citep{Ochmann2024}. 

The X-ray spectrum of NGC 1566, measured in rapid X-ray follow-up observations carried out with XMM-Newton, NuSTAR and Swift \citep{Parker2019}, is a typical type 1 X-ray spectrum at high state, extending to high energies 
and modified by ionized absorption and mild reflection. With the reflection grating spectrometer (RGS) an outflow with velocity of 500 km/s, possibly launched during the outburst, was detected. The amount of X-ray cold absorption does not change between low-state and high-state, implying that variable absorption is not the cause of the outburst. \citet{Parker2019} concluded that a heating front propagating through the disk caused by the Hydrogen instability provides a consistent explanation of the 2018 outburst. 

\begin{figure*}[]
  \centering
\includegraphics[clip, trim=1.7cm 16.5cm 1.0cm 3.3cm, width=11.5cm]{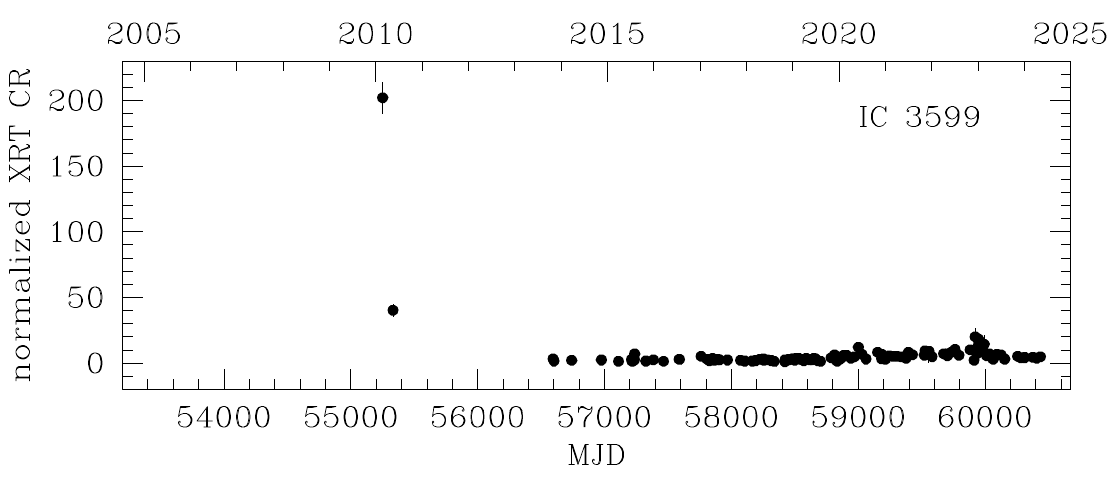}

\includegraphics[clip, trim=1.7cm 16.5cm 1.0cm 3.3cm, width=11.5cm]{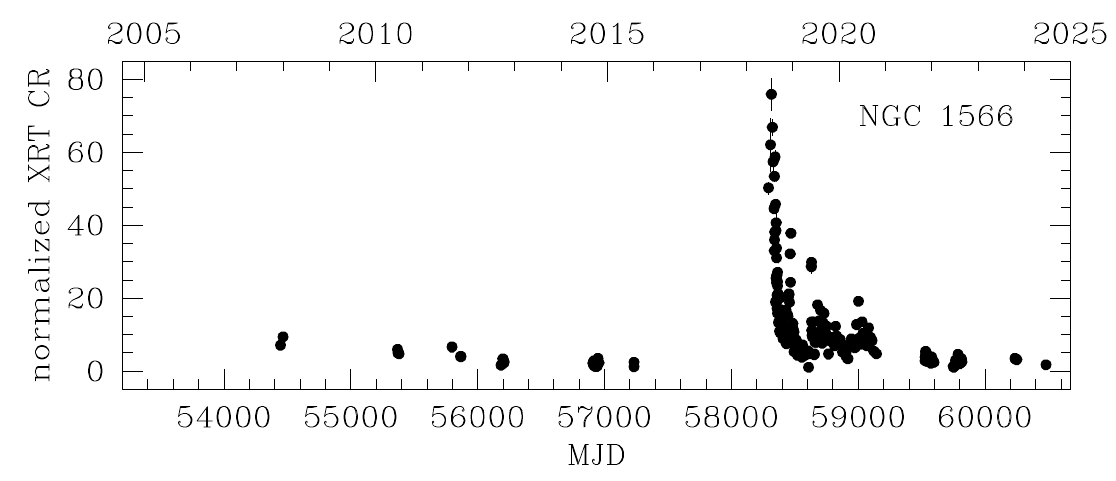}

\includegraphics[clip, trim=1.7cm 16.5cm 1.0cm 3.3cm, width=11.5cm]{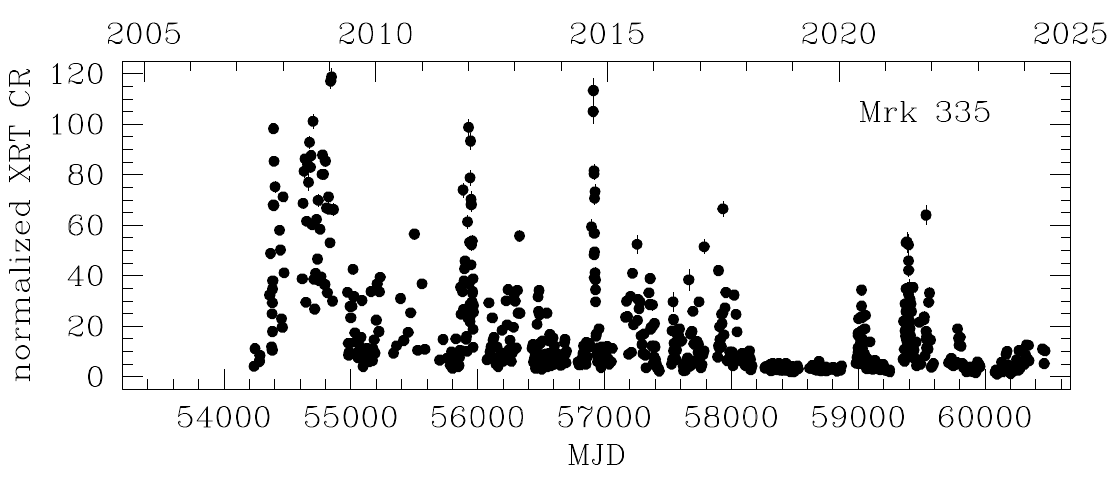}
   \caption{Swift XRT count rate (CR) light curves of highly variable AGN between 2005 and 2024. From top to bottom: IC 3599, NGC 1566, and Mrk 335. For display purposes, X-ray count rates were re-scaled such that the lowest detected count rate was set to 1. This visualizes the total amplitude of the variability and allows a direct comparison between different light curves. 
}
\label{fig:lc_swift}%
    \end{figure*}

An earlier hard X-ray flare is present in the Swift BAT light curve \citep{Oh2018} in 2010, but was not followed up spectroscopically.
The Swift XRT long-term X-ray light curve of NGC 1566, based on our own monitoring observations and additional archival data, is displayed in Fig. \ref{fig:lc_swift}. The X-rays continue to be at low emission levels in 2024. 

NGC 1566 is the nearest of the CL cases detected so far in NLS1 galaxies \citep[see Tab. 1 of][]{Xu2024}, and among the nearest repeating CL AGN known. 

\subsection{Regular intermediate-amplitude dimmings and brightenings of reverberation-mapped AGN and the cases of NGC 5548 and NGC 3516 }
  
\subsection{NGC 5548}
As a nearby, bright, intermediate-type Seyfert galaxy, NGC 5548 has been the target of several reverberation mapping campaigns and many optical spectroscopic studies covering several decades \citep[e.g.,][and references therein]{Denney2010, Bon2016, GardnerDone2017, Horne2021, Lu2022}.  It is variable in the optical and X-ray regime, densely monitored with Swift \citep{McHardy2014, Mehdipour2022}. In low-states its BLR becomes faint, but 
H$\alpha$ and H$\beta$ are always detected. NGC 5548 can therefore be regarded as a CL AGN, changing between Seyfert type 1.5 and type 1.8. 
In the same sense, many reverberation-mapped AGN would represent CL AGN, since pre-selected to show significant continuum variability. 

\begin{figure*}[h]
  \centering
\includegraphics[clip, trim=1.7cm 16.5cm 1.0cm 3.3cm, width=11.5cm]{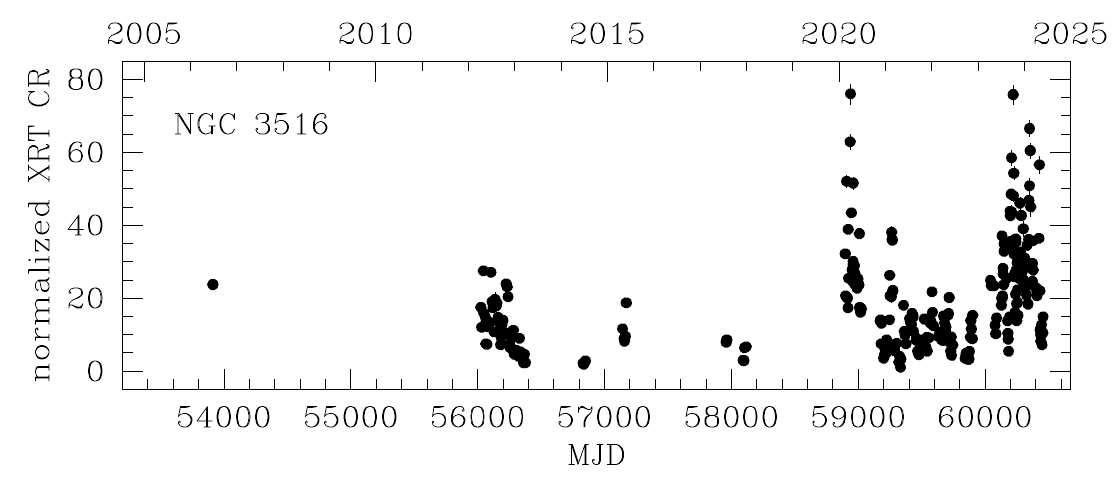}

\includegraphics[clip, trim=1.7cm 16.5cm 1.0cm 3.3cm, width=11.5cm]{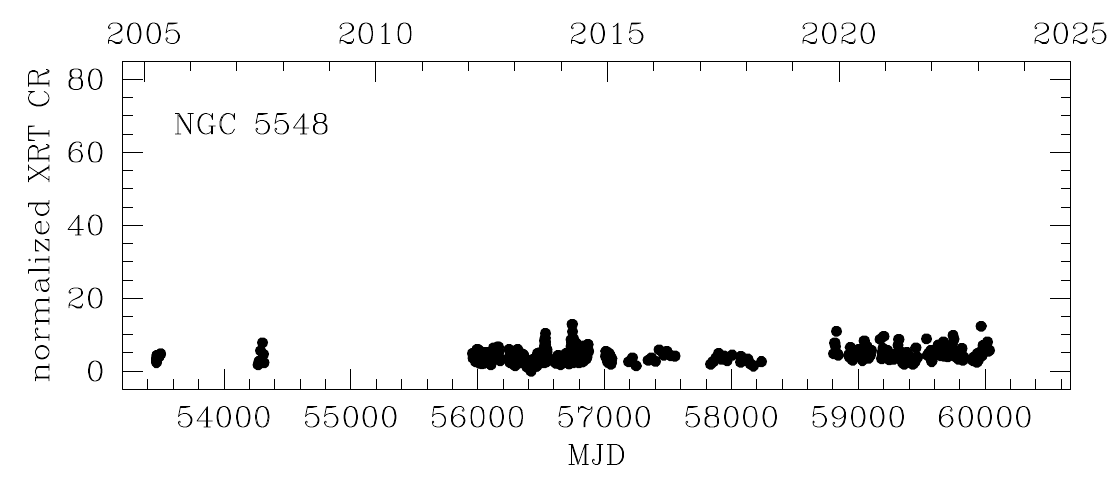}

\includegraphics[clip, trim=1.7cm 16.5cm 1.0cm 3.3cm, width=11.5cm]{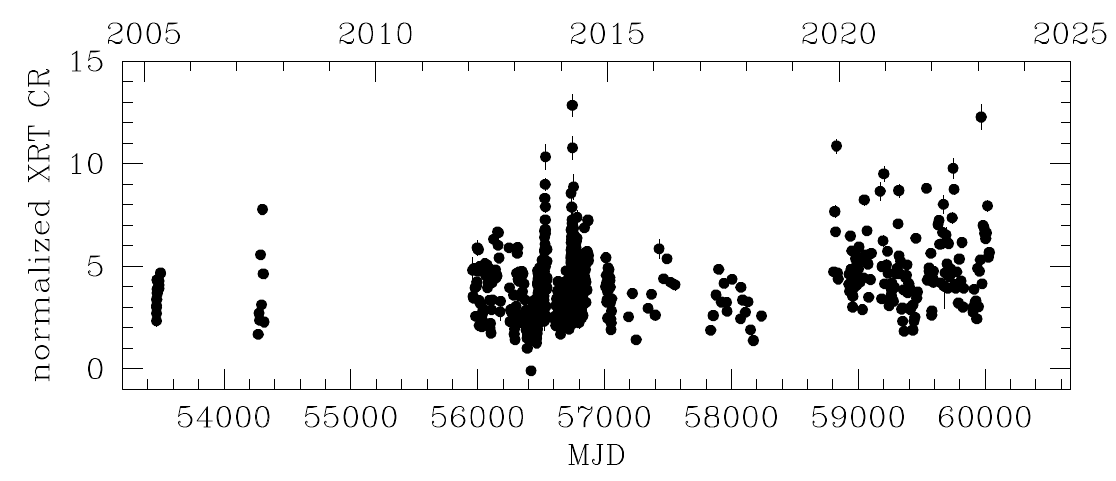}
   \caption{Swift light curves of highly variable AGN between 2005 and 2024, as before in Fig. \ref{fig:lc_swift}, but now from top to bottom: NGC 3516, NGC 5548 (same CR axis scale as the other AGN), and again NGC 5548 (CR axis scale optimally resolved). 
}
\label{fig:lc_swift_02}%
    \end{figure*}

The  amplitude of variability of NGC 5548 in X-rays is much lower than in the other AGN discussed in this contribution (Fig. \ref{fig:lc_swift_02}, based on all archival Swift XRT observations of NGC 5548).
Based on evidence for periodicity of the long-term optical light curve between 1972 and 2015 and accompanying changes in the broad lines with a period of 5700 d, it was suggested that NGC 5548 may host a compact binary SMBH \citep{Bon2016}.

\subsection{NGC 3516}

The Seyfert galaxy NGC 3516 has long been known to be highly variable in its continuum and broad emission lines. Part of the initial sample of \citet{Seyfert1943}, strong broad-line variability was first reported by \citet{Andrillat1968}. 
NGC 3516 stands out as highly variable in the X-ray band as well. For instance, it was one of the most highly variable AGN in a dedicated search for such AGN in the ROSAT data base \citep{KomossaBade1999}.  
Its long-term Swift XRT light curve \citep{Mehdipour2022-n3516}, demonstrating the high-amplitude X-ray variability of this AGN in recent years, based on data retrieved from the Swift archive, is shown in Fig. \ref{fig:lc_swift_02} in comparison with the other AGN discussed in this contribution.  

NGC 3516 has been the target of many spectroscopic monitoring campaigns spanning decades  including reverberation mapping   \citep[e.g.,][and references therein]{Denney2010, Shapovalova2019, Ilic2020, Oknyansky2021, Popovic2023, Ilic2023}, showing that NGC 3516 is a CL AGN which varies between type 1 and type 1.8 or even 1.9 at selected epochs.  
 Based on a detailed study of the H$\beta$ flux and profile variability, \citet{Popovic2023} concluded that an intrinsic mechanism, rather than large-scale obscuration in the outer BLR, is required to explain the CL events in NGC 3516, even though a contribution of dust within the BLR may still play a role at low activity phases. 

\subsection{Frozen-look AGN and the case of Mrk 335}

The NLS1 galaxy Mrk 335 was once among the X-ray brightest AGN. However, when first observed with Swift, its X-ray flux had strongly declined (Fig. \ref{fig:lc_swift}). When followed up with Swift in dense, long-lasting monitoring campaigns \citep{Grupe2007, Gallo2018-m335, Komossa2020-m335}, Mrk 335 did exhibit episodes of high-amplitude flaring back into high-states. However, these episodes became less frequent, less long-lasting and less bright, and since 2018, Mrk 335 has remained in deep X-ray low-states most of the time (Fig. \ref{fig:lc_swift}). This is still the case in our most recent observations from 2024.

Since bright and nearby, Mrk 335 still provides us with good-quality X-ray spectra even in low-state and this is the reason why it was observed multiple times in deep spectroscopic follow-ups, e.g. with XMM-Newton and NuSTAR \citep{Grupe2008, Grupe2012, Gallo2013, Gallo2019, Parker2019-m335, Ezi2021, Kara2023}. However, two very different spectral models still provided successful fits to the complex X-ray low-state spectrum, with no way to distinguish based on the quality of the spectral fit alone: The first one consists of an absorption model where the absorber partially covers our line-of-sight to the continuum source (the accretion disk),  the second one is a reflection model where the intrinsic photons are subject to blurred reflection from the inner accretion disk.  

The solution came from combining X-ray with optical--UV data. This approach showed that X-ray and optical--UV emission are uncorrelated most of the time \citep{Gallo2018-m335, Komossa2020-m335} and that the high-ionization, broad HeII emission line did not follow the simultaneous high-amplitude X-ray variability even though its ionization potential is in the soft X-ray regime \citep{Komossa2020-m335}.
These results can be understood if the X-ray variability is predominantly caused by a (partial-covering, dust-free) absorber located along our line-of-sight, but not covering a full 4$\pi$ sphere shielding the BLR. 
The presence of line-of-sight absorption is independently detected with HST \citep{Longinotti2019} and with XMM RGS \citep{Parker2019-m335} spectroscopy of Mrk 335, revealing UV and X-ray absorption lines.  This result does not exclude a reflection contribution to the X-ray spectrum, but requires a strong absorption component, which drives at least a large fraction of the observed variability. 

In summary, Mrk 335 is a good example for (dust-free) absorption-driven X-ray continuum variability. It is a frozen-look AGN as far as its broad HeII emission is concerned \citep{Komossa2020-m335}, but its Balmer lines still do show a response to the mild optical continuum variability during the epoch it was the target of reverberation mapping \citep{Grier2012}.  Swift monitoring of Mrk 335 is still ongoing in order to trace the duration of the deep low-state, follow the decoupling between X-ray and UV--optical continuum variability, and catch its rise back to high-state. 

\section{Apparent CL variability and impostors}

Not all observed cases of CL events are real. Some observational setups can mimic false BLR variability, instrumental sensitivity needs to be taken into account when assessing the presence or absence of faint Balmer-line emission, and some phenomena look spectroscopically similar to CL AGN but are due to very different physical processes. We discuss these in turn. 

\subsection{Artificial variability, biases and challenges}  

(1) Slit or fiber not on the nucleus. 
Nearby galaxies appear largely extended on the sky, occasionally making it difficult to position a (narrow) slit or fiber exactly on the nucleus. Since BLR and continuum are pointlike (only broadened by atmospheric seeing) whereas the NLR is extended, a mis-positioned slit will provide a spectrum with apparently faint BLR and continuum. That way, dimming events and turn-off CLs can be mimicked.  

(2) Telescope sensitivity. Faint broad Balmer lines will increasingly merge with the noise, and it then depends on the exposure time and telescope sensitivity, whether faint H$\beta$ and/or H$\alpha$ can still be detected in a given spectrum. Accordingly, the turn-off state will be classified as type 1.8, type 1.9, or type 2, even if both Balmer lines are still intrinsically present. It is possible that some of the historic CL events reported to reach type 2 states in the 1970s and 1980s are due to this effect. Complete turn-offs into type 2 states have been rarely observed in recent years. 

\subsection{CL look-alikes: TDEs and SNe (type IIn and/or in dense medium)}

The optical spectra of both, TDEs and supernovae (SNe) IIn can show temporary broad (and/or narrow) optical emission lines, and will thus mimic a changing-look AGN. In fact, there is a parameter space, where all three outburst types (AGN, SNe, and TDEs)  look alike, and are challenging to distinguish from each other.  Here, we describe three scenarios, and discuss ways how to distinguish them from bona fide CL AGN variability \cite[see][for a rigorous discussion of this topic]{Komossa2009}.{\footnote{Further, there is a new class of  long-duration ambiguous transients \citep{Wiseman2024}, even though most or all of them are characteristically different from CL AGN.}} 

(1) SN of type IIn: These type of SNe are known to show temporary broad and narrow emission lines in their optical spectra which look very similar to AGN \citep[][]{Schlegel1990}. The narrow lines arise from strong interactions of the SN ejecta
with dense circum-stellar matter. Strong ionizing radiation may originate from the
shock break-out, Compton-upscattering, as well as in shocks when the SN ejecta collide with the precursor wind and/or the interstellar medium.

(2) SN in dense medium, in a quiescent galaxy or an AGN: Supernovae in a dense medium would particulary efficiently reprocess continuum into emission lines, and can even produce temporary low-level hard X-ray emission (e.g., SNe in the outskirts of the dense torus environment could be partially obscured or extincted, and their decline light curves and continuum emission would look different from other SNe). The situation is especially complicated, if the SN does not happen in a quiescent galaxy but in a long-lived AGN, and all of the SN features would superpose on the long-lived AGN spectrum, which can itself be (mildly) variable.

(3a) TDE in a quiescent host or in an AGN. TDEs are most reliably identified in quiescent galaxies (i.e., systems with dormant SMBHs without a long-lived accretion disk and therefore without a NLR), because TDEs are then the only known mechanism to produce giant X-ray flares of quasar-like peak luminosity \citep{Rees1988, KomossaBade1999}.
In such a scenario, broad emission lines can arise from the temporary accretion disk formed from the disrupted star, and the flung-out stellar debris. If the environment is gas-rich, narrow emission lines can be excited as well, first detected from gas closest to the line of sight (little light travel-time delays).
If a TDE happens in an AGN, all the TDE features overlapp with the long-lived AGN features. 

How can we distinguish all of these impostors from true CL AGN ? One crucial parameter is the X-ray luminosity and X-ray light curve. Another key observation  is repeat optical spectroscopic coverage, ideally including a spectrum {\em{before}} the outburst.
The X-rays are important, because longer-duration (weeks to months) X-rays exceeding ~10$^{41-42}$ erg/s have never been observed from SNe, thus allowing to exclude an SN interpretation of the spectrum.  
Pre-outburst optical spectroscopy is important, because it tells if the host is active or quiescent, based on the presence or absence of a NLR. Similarly, pre-outburst X-ray detections imply a long-lived AGN if $L_{\rm x} > 10^{42}$ erg/s and exclude a TDE in a quiescent host galaxy. 
Finally, post-outburst spectroscopy will tell, if the narrow lines fade (SN IIn or TDE) or are permanent.
The most challenging situation is a TDE or SN in an AGN, in which case the long-term spectroscopic evolution of multiple emission lines provides the best diagnostics to distinguish between all three scenarios. Nevertheless, individual sources could be challenging to classify. PS16dtm (SN\,2016ezh) is a good example. It shows properties of all three, a superluminous SN, a TDE, and a CL NLS1, and all three classifications have been proposed and discussed in the literature \citep{Terreran2016, Dong2016, Blanchard2017, Petrushevska2023}, (Ili\'c, these proceedings).

Finally, TDEs are extremely rare, therefore in a statistical sense we know that at most a tiny fraction of CL AGN  could be mimicked by TDEs, because the observed frequency of CL events in nearby galaxies \citep{Runco2016} is far too high.

\section{(Semi)-periodic variability, candidate binary SMBHs, and the case of OJ 287}

  \begin{figure*}
   \centering
  \includegraphics[clip, trim=0.9cm 5.3cm 1.0cm 1.3cm, width=14.0cm]{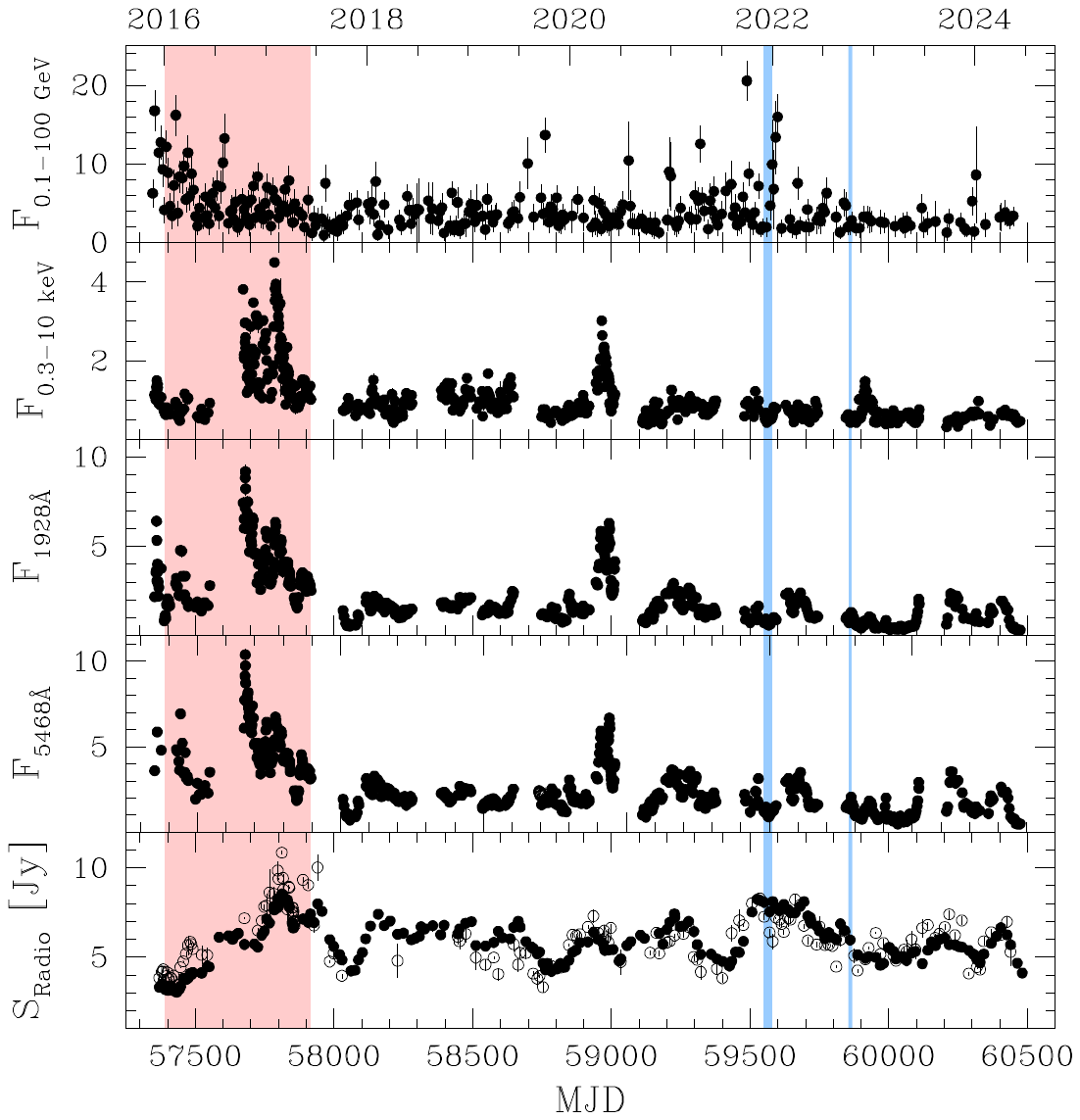} 
   \caption{MWL MOMO light curve of OJ 287 between 2015 and 2024 [from top to bottom: Fermi gamma-ray flux (the highest data point from 2015 is off the plot, and only detections are shown, no upper limits), Swift (0.3--10 keV) X-ray flux, Swift UVOT UV W2 flux at 1928A, Swift V flux at 5468A, and Effelsberg radio flux densities at  10.45 GHz (filled circles) and at 32--36.25 GHz (open circles)]. Fluxes are reported in units of 10$^{-11}$ erg/cm$^2$/s. 
   Two short, bright, gamma-ray flares were detected by Fermi \citep{Abdollahi2023} in late 2021, first discussed as such by \citet{Komossa2022}, overlapping in time with the bright long-lasting radio flare in 2022--2023 and therefore likely related \citep[see the discussion by][]{Komossa2022}, and non-thermal in nature. 
   The epochs of two {\em{optical}} (thermal) outbursts predicted by one of the binary SMBH models (a precursor flare in 2021 and a main outburst in 2022) are marked in blue. No optical outbursts were detected at these epochs. Rather, the
   last optical double-peaked outburst was observed in 2016--2017, marked in red. }
\label{fig:lc_oj287}%
    \end{figure*}

\subsection{Candidate binary SMBH identification from light curves}

Supermassive binary SMBHs form in the course of galaxy mergers \citep{Begelman1980}. The evolved binary SMBHs with small separations, of much less than a parsec, can no longer be spatially resolved with current techniques, and we rely on indirect methods to detect them. A large variety of different detection methods has been explored in the literature \citep[see, e.g.,][for recent reviews]{KomossaZ2016, deRosa2019, dOrazioCharisi2023}. The most common one makes use of (semi-)periodic signals in long-term MWL light curves. These can be induced as a consequence of orbital motion of one or both SMBHs, or precession of the accretion disk or jet. 
At least several periods should be covered in any light curve to be considered a reliable periodicity candidate; otherwise noise can mimic the periodic signal \citep{Vaughan2016}.  

Many recent searches for binary SMBH candidates have been conducted on ongoing optical photometric surveys such as the Catalina Real-time Transient Survey \citep[CRTS;][]{Graham2015a}, the Palomar Transient Factory \citep[PTF;][]{Charisi2016}, and Pan-STARRS \citep{Liu2016}. Recent examples of systems with sinusoidal light curve variations include PG 1302--102 \citep{Graham2015b, dOrazio2015, Kovacevic2019}, and PKS 2131--021 \citep{ONeill2022} (see Sect. 5.3 for the case of NGC 5548, and \citet{Popovic2021} for a recent discussion how to search for binary SMBHs via BLR line profiles). 

A different semi-periodic light curve pattern, not sinusoidal, is shown by the candidate binary SMBH OJ 287. Its long-term optical light curve exhibits recurrent bright double-peaks repeating every 11$\pm$1 yrs, most clearly present in the 1970--1990 light curve of this blazar \citep[see Fig. 1 of][]{Valtaoja2000}. The most recent double-peak has been identified in 2016--2017 \citep{Komossa2023-apj}.
OJ 287 has the most densely covered MWL lightcurve of all binary SMBH candidates \citep[][]{Komossa2021-apj} and is discussed in more detail in the next Section. 

\subsection{OJ 287 and constraints on binary SMBH models from the MOMO project}

Several different binary SMBH scenarios for OJ 287 have been explored in the literature; involving either disk or jet precession, orbital motion of one or two jet-emitting SMBHs, disk impacts of the secondary SMBH during its orbit around the primary, or the presence of mini-disks around two SMBHs fed by a circum-binary disk \citep{Sillanpaa1988, Lehto1996, Katz1997, Villata1998, Valtaoja2000, LiuWu2002, Tanaka2013}. 
According to other suggestions, a binary is not required, or its absence is preferred \citep{Villforth2010, Britzen2018, Butuzova2020}. All of the binary models have offered explanations for the  double-peaked nature of the outbursts and are based on excellent theoretical frame works, but none has so far been able to predict correctly the timing (and spectra) of the outbursts of the recent light curve of OJ 287 since 2000 \citep{Komossa2023-mnras}. 

Tight new model constraints came from ongoing dense MWL monitoring in the course of the MOMO project. It was initiated in 2015 and is designed to distinguish between different binary SMBH scenario predictions, and it independently sheds new light on blazar disk-jet physics
\citep{Komossa2023-an}. Dense monitoring of OJ 287 at a cadence of a day to weeks at multiple frequencies in the radio, optical, UV, X-ray and gamma-ray bands, along with dedicated deeper MWL follow-up spectroscopy at selected epochs, provides timing, spectra, and broadband SEDs at all activity states of OJ 287 \citep[e.g.,][]{Komossa2020, Komossa2021-xmm, Komossa2021-apj, Komossa2022, Komossa2023-apj}. This is the densest and longest MWL monitoring so far carried of OJ 287 at $>$ 12 frequencies, and among the densest of any blazar (Fig. \ref{fig:lc_oj287}). 

Recent MOMO results \citep{Komossa2023-apj, Komossa2023-mnras} have clearly established that (1) no outburst occurred in October 2022 \citep{Valtonen2022} 
and instead the last double-peaked outburst was observed in 2016--2017 (our Fig. \ref{fig:lc_oj287}); (2) the observed upper limit on the accretion-disk luminosity of OJ 287 (based on 3 different estimates) is a factor $>10-100$ fainter than required by a $\sim10^{10}$ M$_\odot$ SMBH accreting at 10\% Eddington; 
instead 
(3) the (primary) SMBH mass derived from BLR scaling relations is modest and of order 10$^8$ M$_\odot$ consistent with the faintness {\citep{Nilsson2020}} of the host galaxy;  and (4) a thermal Bremsstrahlung spectrum with a predicted spectral index of $\alpha_{\nu,opt}$ $\approx -0.2$ has never been observed in the densely covered Swift lightcurve between 2016--2024.

The identification of the last double-peaked outburst in 2016 -- 2017 implies that these outbursts are relatively regular spaced at 11$\pm$1 yr intervals, but are not exactly periodic. The low (primary) SMBH mass implies that OJ 287 is no longer a candidate for detection by near-future pulsar timing arrays. Instead, it would be in the sensitivity regime of LISA-type space-based gravitational-wave interferometers \citep{Colpi2019} upon its final coalescence. 

In summary, successful future binary SMBH models of OJ 287 should reproduce a regular double peak every $\sim$ 11 yr but allow for $\pm$1 yr variations, should be based on a moderate primary SMBH mass of $\sim$10$^{8}$ M$_\odot$ (similar to the approach previously followed by \citet{LiuWu2002}, but now without assuming strict periodicity), and should reproduce synchrotron outbursts. One  alternative binary modelling avenue to follow will be  the application of recent binary merger simulations which predict SMBH binaries with mini-disks fed by a larger circum-binary disk \citep{dOrazio2013}. \citet{Tanaka2013} loosely suggested that such a scenario could apply to OJ 287, but modelling of OJ 287 within that framework has not yet been carried out.   

The long-term light curve of OJ 287 at representative MOMO frequencies is shown in Fig. \ref{fig:lc_oj287}, including previously unpublished X-ray, optical, UV, and radio measurements from 2023--2024. It shows that OJ 287 is at a deep low-state in June 2024, particularly obvious in the optical, UV, and radio bands.{\footnote{A previous deep optical  low-state of OJ 287 was once speculated to possibly represent an epoch in the binary SMBH orbit evolution where the jet of the primary SMBH is briefly misaligned due to the perturbation of a secondary SMBH when it passes near the jet \citep{Takalo1990}, but the deep low-states have happened too often in recent years \citep{Komossa2021-apj, Komossa2023-apj} to be consistent with such an interpretation.}}  
The next double-peak is expected in 2026--2028 \citep{Komossa2023-apj}, and dense monitoring in the course of the MOMO project is planned.

\section{Upcoming time domain surveys in the optical and X-ray regime}

Upcoming time domain surveys in the optical and X-ray bands will not only detect large numbers of extremely variable AGN and CL AGN, but will also facilitate a search for the characteristic periodic signals of binary SMBHs based on the high-cadence light curves coverage every one to few days. In the optical bands, the Vera C. Rubin Observatory Legacy Survey of Space and Time  \citep[LSST;][]{Ivezic2019,Biancho2022} is scheduled to start operation in the year 2025.
LSST will monitor the entire Southern sky for 10 years in 6 filters (ugrizy) with a 9.6 deg$^2$ field-of-view and a 5 $\sigma$ g-band magnitude depth of $\sim$24.5 in a single 30s exposure {\footnote{\url{https://www.lsst.org/scientists/keynumbers}}}. 
LSST will detect millions of variability events per night, issuing an alert within 60s from the detection 
to enable prompt follow-ups with other instruments \citep{Graham2024-lsst}.
The combination of joint photometry data sets from LSST and spectroscopy from other facilities will enable accurate classification for a plethora of events, including outbursts, flares and changing states in AGN. The extent of the sample will further increase the possibility to detect rare events, while the implementation of the Difference Image Analysis \citep[DIA;][]{AlardLupton1998} on the entire dataset will allow us to identify variability in obscured or host-dominated AGN.
In the X-ray regime, two new transient missions will carry out high-cadence all-sky surveys. The Einstein Probe \citep{Yuan2016} was launched in January 2024 and the Space Variable Objects Monitor \citep[SVOM;][] {Wei2016}) was launched in June 2024. 
Several studies have already highlighted how LSST is expected to produce important results in relation to accretion disc studies, reverberation mapping campaigns, and binary SMBH candidate detections in AGN \citep[e.g.,][]{Ivezic2019, Kovacevic2021, XinHaiman2021, Kovacevic2022, Biancho2022}. 
In particular, binary SMBH candidates with periods of months to years, similar to candidates selected from ongoing optical surveys (Sect. 7), are expected to be detected in large numbers in radio-loud and radio-quiet AGN with LSST (longer periods) and in X-rays. 
Further, an advantage of LSST is its high cadence{\footnote{the exact cadence is still being discussed and may change during its 10 year mission; the higher it is, the more short-period systems will be well covered}} so that it can find shorter periods, and it is very deep, therefore it covers AGN with lower-mass SMBHs, of which there are many more, and also they are longer-lived at periods of weeks compared to more massive SMBHs.    This implies that the LSST catalog will contain large numbers of fainter short-period binaries.

Evolved, ultra-compact SMBH binaries of much shorter periods of only days which then coalesce years later in the LISA time window
can be detected in the optical band with LSST through the periodic light-curve modulation from the circum-binary disk feeding the two SMBH's mini-disks \citep{XinHaiman2024},
while the orbital modulation from the two mini-disks themselves is best searched for in the X-ray regime, given the compactness of the mini-disks at very advanced stages of merging. 

Other future surveys with a strong time-domain component and expected to start by the late 2020s are the Cosmological Advanced Survey Telescope for Optical and UV Research (CASTOR), and the Nancy Grace Roman Space Telescope, in the NIR range. These two spacecrafts will add further MWL coverage to the variable Universe, and will have a strong synergy with contemporary surveys such as LSST and the EUCLID satellite (even though the latter does not have an explicit time-domain focus).
%




\section*{Acknowledgments}
 We would like to thank the Swift team for carrying out the observations we proposed.
 In addition to our own data, we have also used the Swift archive at \url{https://swift.gsfc.nasa.gov/archive/}. 
 We would like to thank all participants of the ``VI. Conference on Active Galactic Nuclei \& Gravitational Lensing'' (Serbia, June 2024) for enlightening discussions. S.K. and Z.H. would also like to thank the participants of the conference on ``Multi-messenger observations of supermassive black hole binaries'' which was supported by the Lorentz Center (Leiden, June 2024) for very stimulating discussions. 
 S.K. would like to thank R. Antonucci, D. Hutsemékers, F. Marin, and M. Ochmann for very useful discussions. 
 E.B., N.B., D.I., A.B.K, J.K.D., S.M.M., and L.\v C.P.  
 acknowledge funding provided by the University of Belgrade—Faculty of Mathematics (contract 451-03-66/2024-03/200104) and/or Astronomical Observatory Belgrade (contract 451-03-66/2024-03/200002), through grants by the Ministry of Education, Science, and Technological Development of the Republic of Serbia. D.I. acknowledges the support of the Alexander von Humboldt Foundation. A.B.K. and L.\v C.P. thank the support by the Chinese Academy of Sciences President’s International Fellowship Initiative (PIFI) for visiting scientists. 
 V.P. and D.D. acknowledge the financial contribution from PRIN-MIUR 2022 and from the Timedomes grant within the ``INAF 2023 Finanziamento della Ricerca Fondamentale''. D.D. also acknowledges PON R\&I 2021, CUP E65F21002880003. 
 Z.H. acknowledges support by NSF grant AST-2006176 and NASA grants 80NSSC22K0822 and 80NSSC24K0440.
 K.K.G. acknowledges the Belgian Federal Science Policy Office (BELSPO) for the provision of financial support in the framework of the PRODEX Programme of the
 European Space Agency (ESA).
 This work is partly based on data obtained with the 100\,m telescope of the Max-Planck-Institut f\"ur Radioastronomie at Effelsberg.   
This work made use of data supplied by the UK Swift Science Data Centre at the University of Leicester \citep{Evans2007}. This work has made use of public Fermi-LAT data \citep{Abdollahi2023}.  
This research has made use of the XRT Data Analysis Software (XRTDAS) developed under the responsibility of the ASI Science Data Center (ASDC), Italy.
This work has also made use of the NASA Astrophysics Data System Abstract Service (ADS), and the NASA/IPAC Extragalactic Database (NED) which is operated by the Jet Propulsion Laboratory, California Institute of Technology, under contract with the National Aeronautics and Space Administration.


\end{document}